\documentclass[aps,
prl,
showpacs,floatfix]{revtex4}
\usepackage{graphicx}
\usepackage{bm}

\begin{document}

\title{Superfluid Fermi liquid in a unitary regime}
\author{L.~P.~Pitaevskii}
\date{\today}
\affiliation{CNR INFM-BEC and Department of Physics, University of Trento, I-38050 Povo,
Trento, Italy; \\
Kapitza Institute for Physical Problems, ul. Kosygina 2, 119334 Moscow,
Russia}

\begin{abstract}
A short popular review of theoretical and experimental results about properties of
ultracold Fermi gases in vicinity of the Feshbach resonance is presented.
The paper is based on the authors talk on the meeting of the Physical
Department of the Russian Academy of Sciences dedicated to 100th anniversary of
the birth of L.~D.~Landau (Moscow, January 22, 2008). A version of this
article has been published in \textit{Physics - Uspekhi} \textbf{51} 603
(2008).
\end{abstract}

\maketitle

%DOI: 10.1070/PU2008v051n06ABEH006548
%DOI: 10.3367/UFNr.0178.200806i.0633

When choosing the subject of my presentation at this session dedicated to
the 100th anniversary of the birth of Landau, I wanted to speak about
something that would have surprised Landau. I believe that the recently
created physical object - a universal superfluid Fermi liquid - meets this
requirement in the best way possible.

As is well known, Landau did not regard the microscopic theory of fluids as
a problem worth being occupied with. I quote a well-known passage from
Statistical Physics (see \cite{LL5}, page 187):

\textquotedblleft Unlike gases and solids, liquids do not allow a
calculation in a general form of the thermodynamic quantities or even of
their dependence on temperature. The reason lies in the existence of a
strong interaction between the molecules of the liquid while at the same
time we do not have the smallness of the vibrations which makes the thermal
motion in solids especially simple. The strength of the interaction between
molecules makes it necessary to know the precise law of interaction in order
to calculate the thermodynamic quantities, and this law is different for
different liquids.\textquotedblright 

This statement is perfectly correct for all liquids existing in nature.
However, progress in experimental techniques has recently enabled preparing
liquids with properties independent of any quantities that characterize the
interaction. This situation emerges because the interatomic interaction in
these bodies is, in a sense, infinitely strong. The case in point is
ultracold gases near the so-called Feshbach resonances.

First of all, one can pose the question: what is the word "liquid" taken to
imply? We accept a natural definition: a liquid is a fluid body with a
strong interaction between its particles. We emphasize that fluidity implies
the absence of strict periodicity, of a crystalline long range order. The
liquids of interests to us are made from gases whose atoms obey the Fermi
statistics. The gas is dilute in the sense that the average interatomic
distance  $n^{-1/3}$, where $n$ is the atomic number density, is much
greater than the characteristic range $r_0$ of interatomic forces: 
\begin{eqnarray}
r_{0} \ll n^{-1/3} \ .   \label{gas}
\end{eqnarray}
Condition (\ref{gas}) is always satisfied for the objects under
consideration. However, the fulfillment of this condition does not yet mean
that we are dealing with a gas in the sense that the interaction is weak.
Let the temperature be sufficiently low, such that the gas is degenerate,  $%
T \leq E_{F}$ \cite{footnote}. It is then valid to say that all properties
of the body depend on one parameter  $f$, the amplitude of the scattering of
atoms with the orbital momentum $l=0$ by each other. The interaction is
weak, i.e. the body is indeed a gas, if the amplitude is small in comparison
with the interatomic distances: 
\begin{eqnarray}
|f| \ll n^{-1/3} \ .  \label{interaction}
\end{eqnarray}
The quantities $r_{0}$ and $|f|$ are typically of the same order of
magnitude and conditions (\ref{gas}) and  (\ref{interaction}) are
practically equivalent. However, this is not the case when a system of two
atoms has an energy level close to zero. According to the general theory,
the scattering amplitude then takes the form (see, for example \cite{LL3})  
\begin{eqnarray}
f(k)=-(a^{-1}+ik)^{-1}\ , \label{f}
\end{eqnarray}
where  $k$ is the wave vector and $a=-f(0)$ is the scattering length, a
constant that characterizes  the scattering completely. When  $a>0$, the
system of two atoms has a bound state with the negative energy $\epsilon =
-\hbar^{2}/ma^2$. When $a<0$, the system is said to have a virtual level. If 
$|a|$ is high enough, $|a| \ge k^{-1} \sim n^{1/3}$, the interaction
weakness condition (\ref{interaction}) is certainly violated and we are by
definition dealing with a liquid, although a dilute liquid in the sense of
condition (\ref{gas}). In this case, the properties of the liquid are
characterized by the sole parameter $a$. When $|a| \gg k^{-1}$, the
scattering amplitude reaches its ``unitary limit'' 
\begin{eqnarray}
f \approx \frac{i}{k} \ .  \label{un}
\end{eqnarray}
The length $a$ then also drops out of the theory and we are dealing with a
universal liquid, which properties do not depend on interaction at all. Of
course, the picture under discussion implies the possibility of changing the
scattering amplitude. This opportunity arises in the presence of Feshbach
resonances, in the vicinity of which the position of the energy level of the
system of two atoms depends on magnetic field  \cite{Fesh}. Here the scattering
length as a function of magnetic field can be represented as 
\begin{eqnarray}
a=a_{bg} \left ( 1-\frac{\Delta_B}{B-B_0} \right ) \ .  \label{Fesh}
\end{eqnarray}
Near the resonance $B \approx B_0$, the scattering length is large and the
system is a universal ''unitary'' liquid.

We qualitatively consider the properties of the system at  $T=0$ in
different ranges of the scattering length  $a$. When this length is positive
and relatively small,  $r_0 \ll a \ll n^{-1/3}$, the system of two atoms has
a bound state and the atoms combine  to form molecules with a binding energy 
$\epsilon$.  The system is a Bose gas of weakly bound diatomic molecules, or
dimers. It is  significant that the dimer-dimer scattering length  $a_{dd}$
is also positive, i.e. these molecules experience mutual repulsion.  Calculating $%
a_{dd}$ is an intricate problem, which was solved by Petrov, Salomon and
Shlyapnikov \cite{PSS04}. It turned out that $a_{dd}=0.6a$. Therefore, in
this regime, the system is a weakly nonideal Bose gas of molecules described
by the Bogoliubov theory\cite{B47}, with the obvious change $m \to 2m,a \to
0.6a$.

The question of the lifetime of this system is of paramount importance for
the entire area of physics involved. This lifetime is limited by transition
from the weakly bound level to deep molecular levels in molecular collisions
accompanied by the release of large amount of energy. The molecular number
loss in these inelastic processes is described by the equation ${\dot{n}_{d}}%
=-\alpha _{dd}n_{d}^{2}$. The dependence of the recombination coefficient $%
\alpha _{dd}$ on $a$ was also investigated in \cite{PSS04}. It turned out
that $\alpha _{dd}\propto a^{-2.25}$. Therefore, the system becomes more
stable with an increase in the scattering length, i.e., as the resonance is
approached. This paradoxical result stems from the Fermi nature of atoms or,
to be more precise, from the fact that fermions with parallel spins cannot
reside at the same point. In a Bose gas, which was also studied in
experiments, the lifetime decreases sharply as the resonance is approached.
This is the reason why only the Fermi liquid can actually be investigated in
the unitary regime. The experimentally measured dependence of $a_{dd} $ on $a$ is
presented in Fig. 1. It is in satisfactory agreement with the theory.

\begin{figure}[t]
\includegraphics[width=7cm,height=7cm]{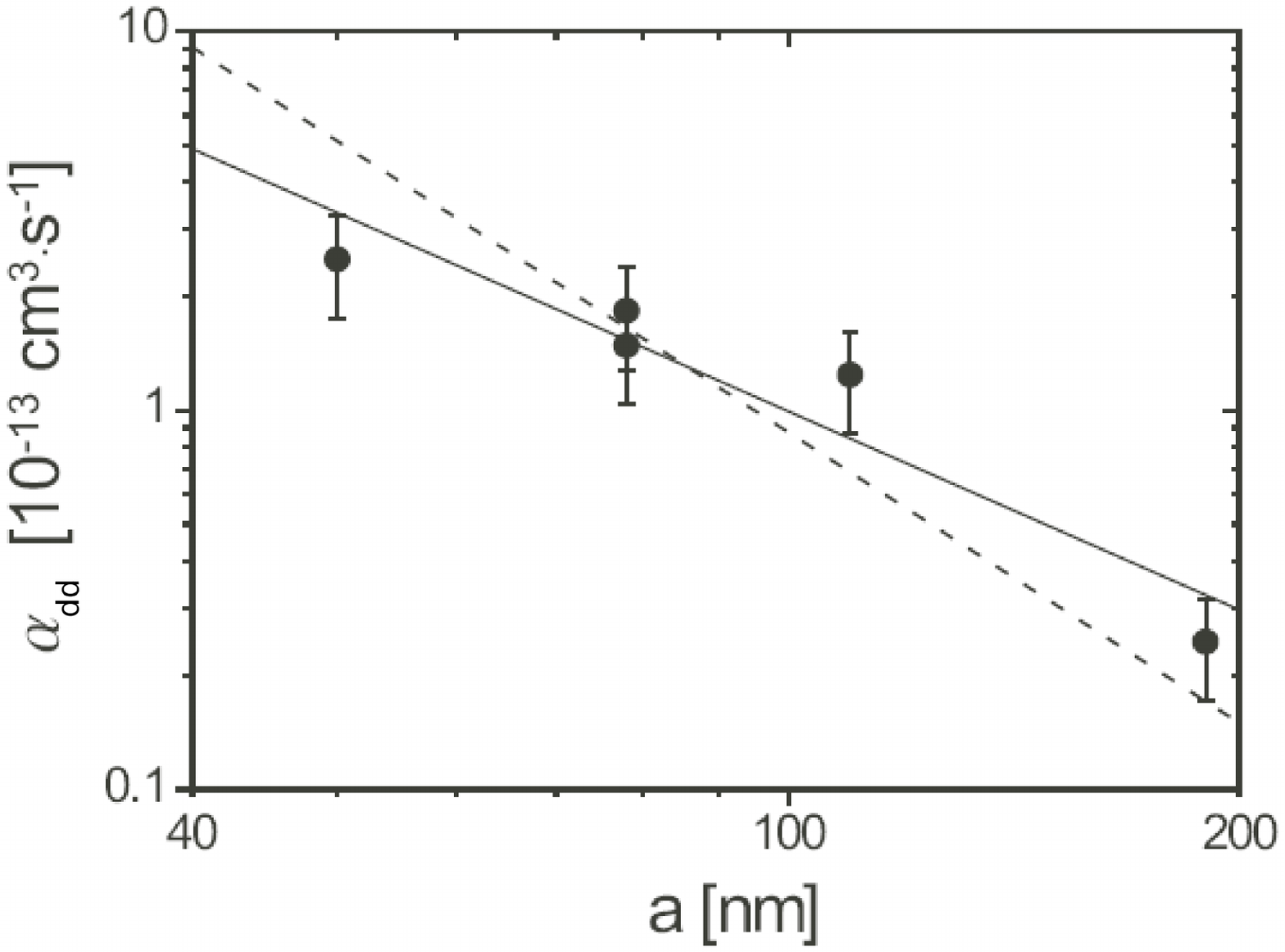}
\caption{The dimer recombination coefficient $a_{dd}$ as a function of the
scattering length $a$ (from \cite{PSS05}). The slope of the dashed
line corresponds to the theoretical dependence $a_{dd} \propto a^{-2.5}$.}
\label{fig1}
\end{figure}

We now consider the opposite limit case, where the scattering length is
negative and small in modulus, $a < 0, r_0 \ll |a| \ll n^{-1/3}$, as is the
case on the opposite side of the resonance. The system is then a weakly
nonideal Fermi gas with attraction between the atoms. According to the
theoretical concepts of Bardeen - Cooper - Schrieffer and Bogoliubov, the
occurrence of a Fermi surface gives rise to Cooper pairs in this case. As a
result, a gap appears in the fermion energy spectrum and the system becomes
superfluid. In the immediate vicinity of the resonance, the system is a
universal unitary Fermi liquid. Because the system is superfluid in both
limit cases considered, it is reasonable to assume that it is superfluid in
all of the interval of a values. (Different arguments are presented below.)
Of course, the system is then assumed to be stable in the unitary regime. This
assumption is supported by the wealth of experimental data and theoretical
calculations.

Prior to discussing these results, I briefly describe the typical
experimental arrangement using the example of a device at Duke University 
\cite{Thomas04} (see Fig. 2). The atoms are confined in an optical trap formed by a
focused laser beam. The chosen light frequency is somewhat lower than the
absorption line frequency, and therefore the atoms are `attracted' to the
intensity peak. Because the intensity near the focus decreases rapidly in
the radial direction and slowly in the axial direction, the sample was
elongated and cigar-shaped. Solenoids induce the magnetic field required to
attain the resonance. Since the main objective of the experiments was to
investigate superfluidity, two kinds of fermions were needed. In
superconductivity theory, electrons with opposite values of spin projection
are usually considered. In our case, atoms in different hyperfine structure
states were used.

\begin{figure}[t]
\includegraphics[width=7cm,height=7cm]{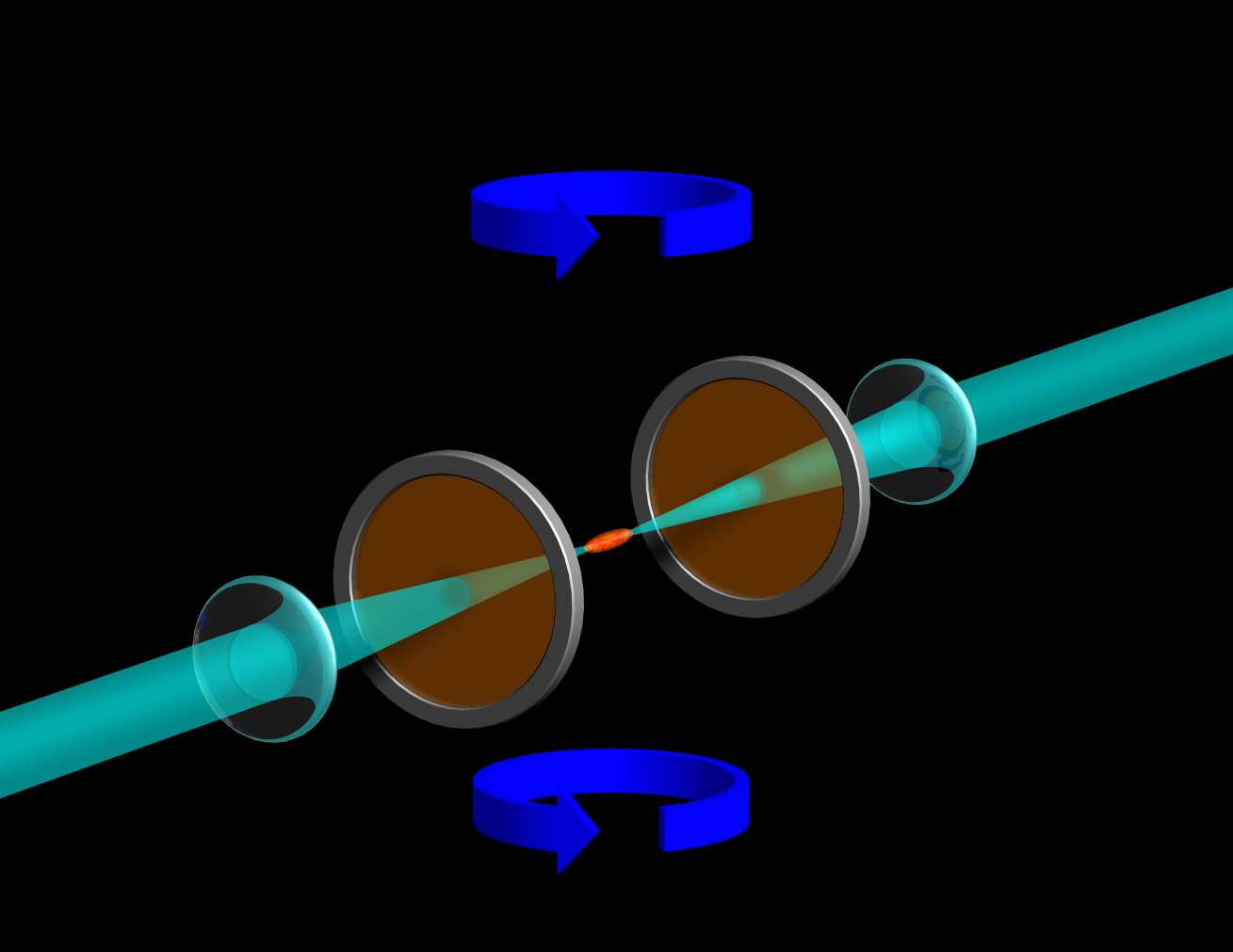}
\caption{Schematic of the device employed at Duke University to
investigate the properties of a Fermi gas in an optical trap near a Feshbach
resonance (from \cite{Thomas04}).}
\label{fig2}
\end{figure}

Experiments with fermions are difficult and the number of groups working
with them is smaller than the number of groups investigating the
Bose-Einstein condensation. The work is undertaken at the JILA (Joint
Research Institute of the National Institute of Standards and Technology and
the University of Colorado) (Boulder), Massachusetts Institute of Technology
(MIT) (Boston), Duke University (Durham), and Rice University (Houston) in
the USA, the \'{E}cole Normale Sup\'{e}rieure (Paris) in France, and the University
of Innsbruck in Austria. It is a pleasure for me to mention that A Turlapov,
one of the leading experimenters at Duke University, has returned to Nizhnii
Novgorod and is making a facility there.

Two types of Fermi atoms were actually used in
the experiments, $^6$Li and $^{40}$K isotopes. The isotope choice was dictated
by the presence of a Feshbach resonance in a convenient range of the
magnetic field and the occurrence of spectral lines in a convenient
wavelength range. 

I give the typical parameters of recent experiments. The number of atoms in
the trap is $N \sim 3 \times 10^{6}-10^{7}$ and the atom density at its
center is $n \sim 2 \times 10^{12}$. Accordingly, the Fermi energy is $E_F
\sim $200-500 nK and the magnitude of the Fermi wave vector is $k_F \sim 0.3
\mu m^{-1}$. The parameters of the trap are conveniently characterized by
the frequencies of atomic oscillations in it. The radial frequency $%
\nu_{\perp}$ normally lies in the 60 - 300 Hz range and the longitudinal
frequency $\nu_z \sim 20$ Hz. The lowest attainable temperature turns out to
be under 0.06$E_F$, i.e., of the order of 10 nK. As is evident from the
subsequent discussion, it has been possible not only to conduct experiments
at these prodigiously low temperatures but also to set up a thermodynamic
temperature scale in this domain. I cannot discuss here the techniques of
gas cooling, and only mention that during the final stage, the gas is cooled
due to the evaporation of the faster atoms from the trap, much like tea is
cooled in a cup left on a table.

One of the most important experimental tasks was to prove that the
system was superfluid. An immanent property of superfluidity is the
existence of quantized vortexes. The velocity circulation around a vortex in
a Fermi liquid is $\Gamma = \pi \hbar/m$, two times smaller than in a Bose
liquid. Accordingly, in the rotation with a sufficiently high angular
velocity $\Omega $, the number of vortexes per unit area must be equal to $%
2\Omega m/(\pi \hbar)$. How can the liquid be set in rotation? MIT
experimenters positioned a pair of thin laser beams along the trap axis,
which were shifted from the axis (Fig. 3) \cite{Kett06}. This `mixer'
rotated about the axis and stirred the liquid. At some instant, the trap was
disengaged, the liquid expanded, and observations of the density
distribution were made. The result is shown in Fig. 4. The vortex cores are
observed as dark reduced-density domains. A simple calculation of the number
of vortexes confirms the theoretical value of the circulation given above.

\begin{figure}[t]
\includegraphics[width=7cm,height=7cm]{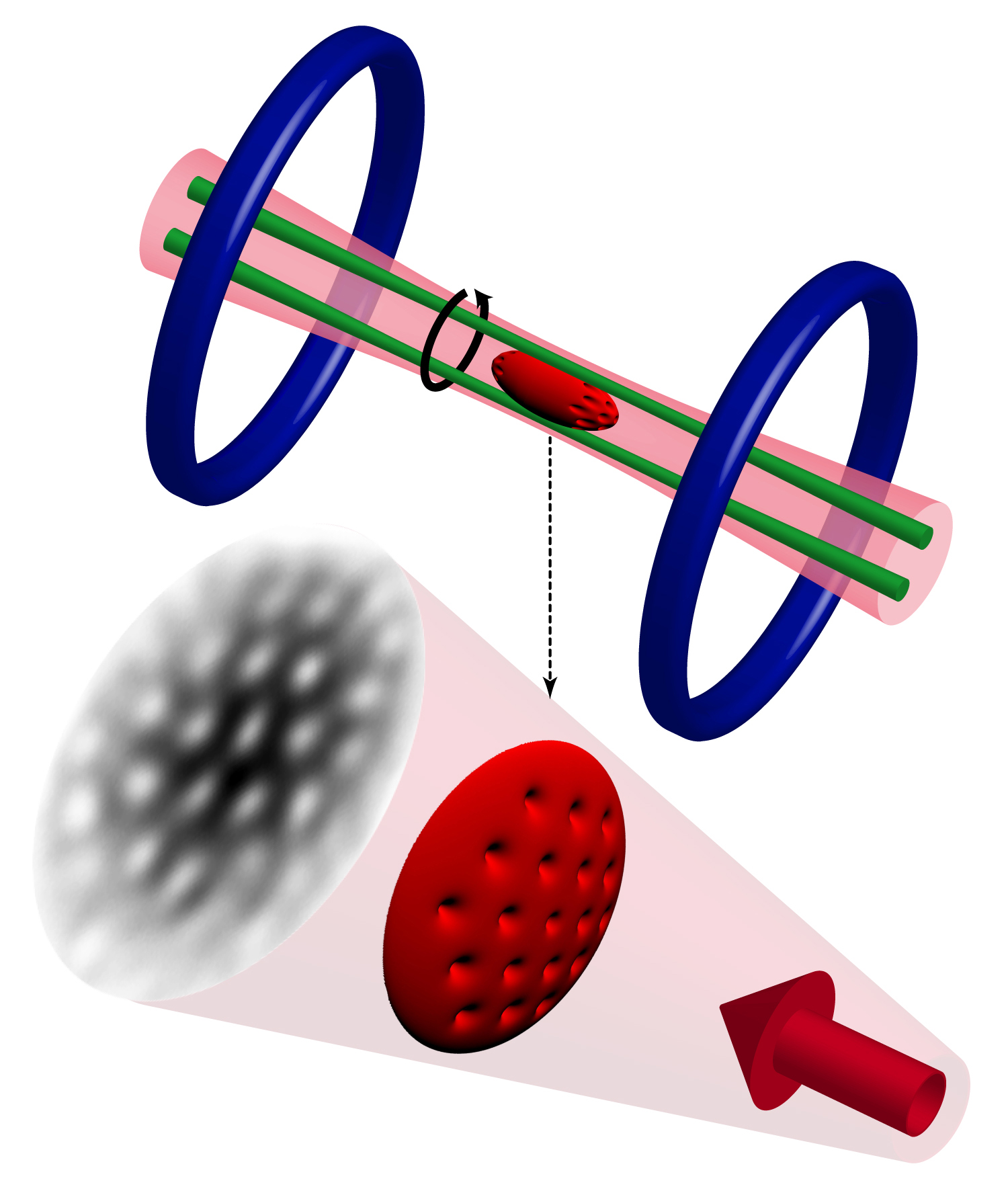}
\caption{Schematic of the device used at MIT for investigating the
rotation of a superfluid Fermi gas in the experiment \cite{Kett06}.
Two laser beams aligned with the axis set the gas in rotation. Separately
shown is the scheme for observing the vortexes from the resonant absorption
imaging of the expanding fermionic cloud.}
\label{fig2b}
\end{figure}

\begin{figure}[t]
\includegraphics[width=8cm,height=4cm]{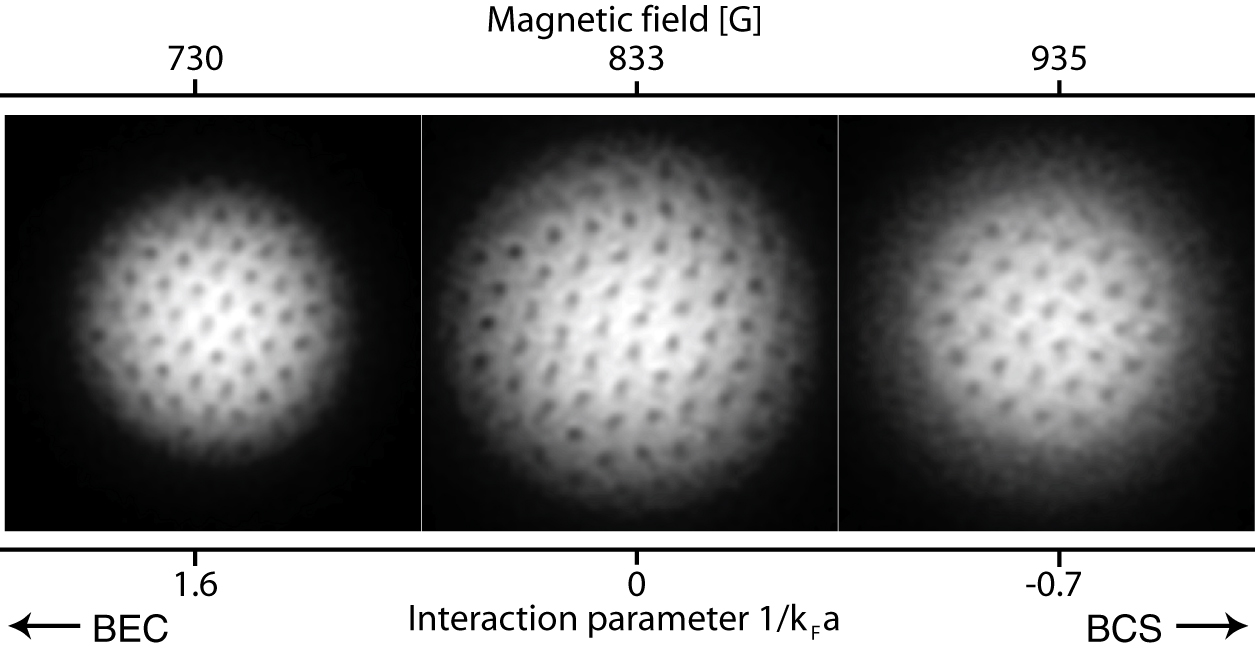}
\caption{ Quantized vortexes in a rotating superfluid Fermi gas (from [8]):
(a) corresponds to a dilute gas of dimers, (b) to a Fermi liquid in the
vicinity of the unitarity point, and (c) to a dilute Fermi gas with a weak
attraction between the atoms.}
\label{fig3}
\end{figure}

We now consider the liquid precisely at the resonance point, when $a \to \pm
\infty $. (It is appropriate to note that this is not a phase transition
point.) We begin from the properties of a uniform liquid at $T = 0$. Apart
from the density, there are no parameters at our disposal on which the
thermodynamic functions may depend. Dimensionality considerations suggest,
e.g., that the chemical potential of the liquid must be of the form 
\begin{eqnarray}
\mu(n)=\xi \mu^{id}(n) \ ,  \label{mu}
\end{eqnarray}
where $\mu^{id}(n)=(3\pi^2n)^{2/3}(\hbar^2/m)$ is the chemical potential of
an ideal Fermi gas with the density $n$ for $T = 0$ and $\xi$ is a
dimensionless coefficient independent of the kind of liquid. The theoretical
task consists in the calculation of $\xi$ and the experimental task involves
its measurement. The first estimates of $\xi$ were made proceeding from
the Bardeen-Cooper-Schrieffer-Bogoliubov (BCSB) theory. This theory is a
mean-field theory and, needless to say, is inapplicable near the unitarity
point. But its ingenious generalization to the strong-coupling case has
allowed obtaining formulas sound in both limit cases (see, e.g., \cite%
{Nozieres85,Randeria95}). At exactly the unitarity point, 
this theory yields $\xi =0.59.$ The most reliable result is provided by 
calculations involving the quantum
Monte Carlo (QMC) technique: $\xi = 0.42$ \cite{Carlson05}. It is noteworthy
that the absence of a small parameter in the theory is substantially
favorable to numerical calculations. The existence of such a parameter often
impairs convergence. An attempt has been made to apply the $\varepsilon$%
-expansion technique, which relies on the fact that $\xi=0 $ in a
four-dimensional space \cite{epsilon}. The theory is constructed in the
space of $D = 4 - \varepsilon $ dimensions under the assumption that $%
\varepsilon$ is small, and the results are then extrapolated to $\varepsilon
= 1$. This technique, which is highly beneficial in the theory of phase
transitions, supposedly yields poor accuracy in this case.

It is significant that the parameter $\xi < 1$. This means that the
interaction at the unitarity point lowers the fluid pressure, i.e., is an
effective attraction. It is therefore reasonable that it leads to fermion
pairing and to superfluidity. A quantitative characteristic of the pairing
is the gap $\Delta $ in the Fermi branch of the spectrum. Once again, the
dimensionality considerations suggest that 
\begin{eqnarray}
\Delta(n)=\theta \mu^{id}(n) \ .  \label{Delta}
\end{eqnarray}
QMC calculations yield $\theta = 0.5$ \cite{Carlson05}.

We now turn our attention to the experimental verification of the theory.
The most direct method of determining $\xi$ consists in the precise measurement
of fluid density in the trap. In the semiclassical approximation, this
distribution is given, in view of expression (\ref{mu}), by the equation $%
\xi \mu^{id}[n(\mathbf{x})]+V(\mathbf{x})=const$. Fitting to the observed
distribution allows determining $\xi$. At Rice University, the value $\xi =
0.46$ was thus found for $^6$Li \cite{Hulet06}. Another method was applied
by experimenters at JILA, who worked with $^{40}$K \cite{Jin06}. They
measured the density distribution and calculated the potential energy $%
U_{pot}= \int n(\mathbf{x})Vn(\mathbf{x})d\mathbf{x}$ of the liquid, which
is proportional to $\sqrt{\xi}$. By this means, they obtained the value $\xi
= 0.46$. The proximity of the values for $^6$Li and $^{40}$K to the
theoretical one confirms the universal nature of $\xi$. Reliable
measurements of the gap $\Delta$, in my opinion, have not been made to the
present day.

Important information about the properties of the liquid may be obtained by
investigating its oscillations in the trap. These oscillations are described
by the Landau superfluid hydrodynamics \cite{Landau41}. (I emphasize that
Landau believed from the outset that his equations applied both to Bose and
Fermi superfluid liquids.) An especially simple result for the oscillation
frequencies in a harmonic trap is obtained for a liquid with the polytropic
equation of state $\mu (n) \propto n^{\gamma}$. We consider an important
type of oscillation: axially symmetric radial oscillations whose frequency
is $\omega = \sqrt{2(\gamma+1)}\omega_{\perp}$\cite{Stringari03}. According
to this formula, in the molecular limit ($a > 0, na^{3} \ll 1$), when $\mu
\propto a_{dd}n$, i.e., $\gamma = 1$, the frequency $\omega=2\omega_{\perp}$%
. In the unitary limit and BCSB limit, $\gamma= 2/3$ as in an ideal Fermi
gas and $\omega = \sqrt{10/3}\omega_{\perp}=1.83\omega_{\perp}$. For
intermediate values of $a$, the frequency cannot be calculated analytically,
but it appears reasonable that the frequency for $a > 0$ is monotonically
decreasing with increasing $a$. These were precisely the indications of the
first experiments. Theories that have this property and rely on the
mean-field approximation have also been proposed. However, the situation is
not that simple. For $na^{3} \ll 1$, the theory permits rigorous
calculations of not only the first term in $\mu $ but also a correction, which
was first determined in \cite{LHY57}. This gives a correction to the
frequency equal to \cite{PS98} 
\begin{eqnarray}
\delta\omega/\omega= +0.72\sqrt{[n(\mathbf{x}=0)a^3_{dd}]} \ .
\label{deltaomega}
\end{eqnarray}
The positive correction sign signifies that the frequency must initially
increase with increasing $a$ and only then decrease to attain the limit value $%
1.83\omega_{\perp}$. This reasoning was disputed on the grounds that
molecular dimers are nevertheless not entirely bosons. However, the correction %
(\ref{deltaomega}) bears a clear physical meaning. It stems from the
contribution to the energy made by zero-point phonon oscillations, whose
occurrence in the superfluid liquid is beyond question. This is why it is
anomalously large, of the order of the square root of the gas parameter $na^{3},$
while the `normal' expansion is performed in this parameter. All this leads
us to the statement that the author has been vigorously promoting, namely,
that the monotonic behavior of the frequency would imply a catastrophe for
the theory. Fortunately, the situation has recently been clarified. New
experiments do yield above $2\omega_{\perp}$ values of the frequency on the
molecular side of the resonance. They are in good agreement with the
calculations throughout the interval of a values performed by the QMC
technique \cite{Astr05} (see Fig. 5). We note that the disagreement with the
data of previous experiments is probably due to the fact that the
temperature in those experiments was not low enough. Meanwhile, correction %
(\ref{deltaomega}) is actually temperature sensitive, because it is related to
the excitation of relatively low energies $\hbar \omega \sim \mu$.

\begin{figure}[t]
\includegraphics[width=8cm,height=4cm]{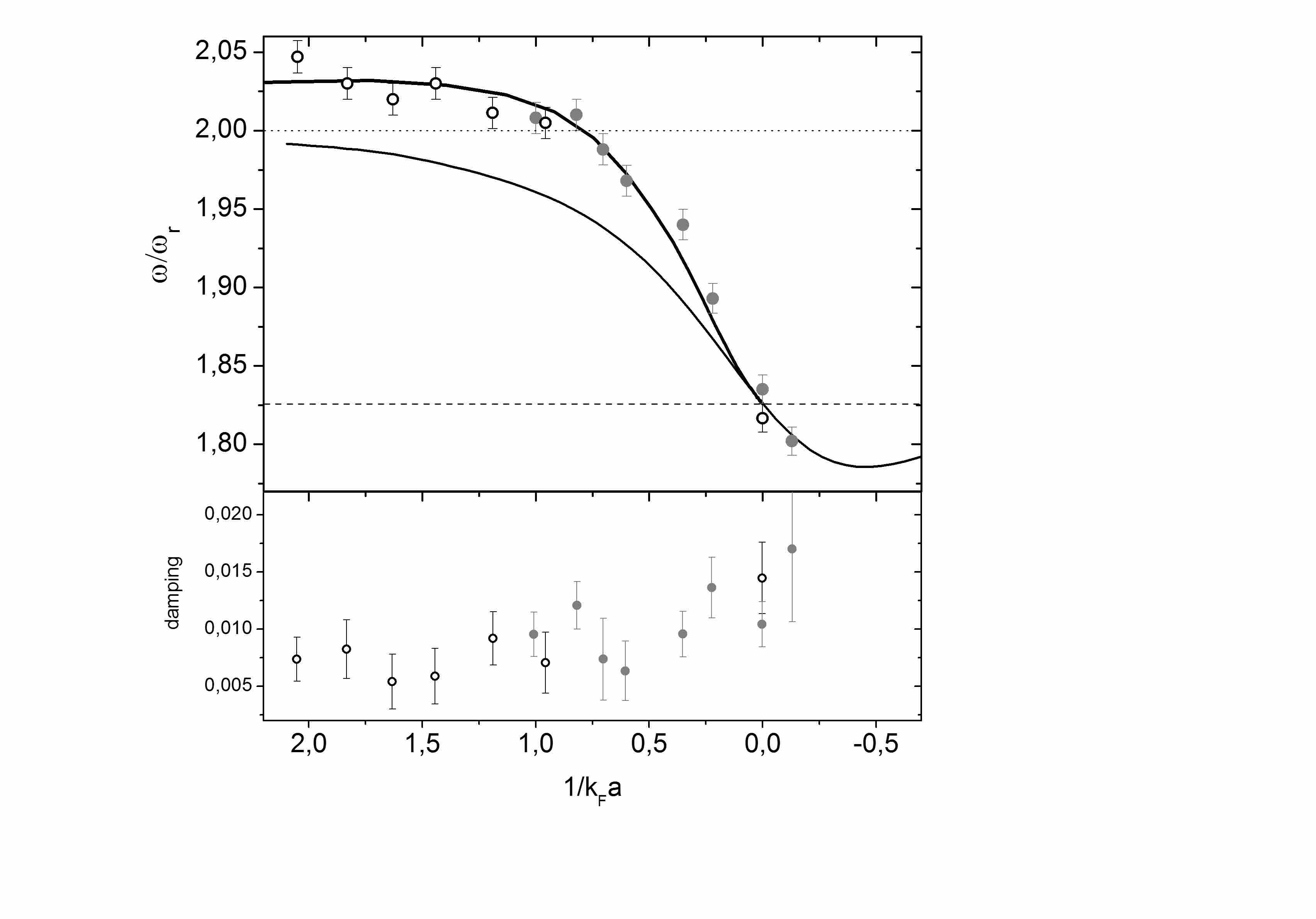}
\caption{ (a) Frequency of radial oscillations as a function of the
scattering length. The upper solid curve represents the data calculated by
the quantum Monte Carlo technique and the lower one is the result of
calculations by the mean-field theory. The points stand for experimental
values. The upper and lower dashed straight lines show the limit frequency
values in a dilute dimer gas and the unitarity point. (b) Measured values
of oscillation damping (from \protect\cite{Grimm07}).}
\label{fig4}
\end{figure}

We now discuss the fluid properties at the unitarity point at finite
temperatures. In this case, the temperature is assumed to be not too high,
and therefore the wavelength of atoms in their thermal motion is long in
comparison with the atomic size: $r_0 \ll \hbar/\sqrt{mT}$. We are actually
dealing with temperatures of the order of $E_F$.

The question of the temperature of transition to the superfluid state is
all-important here. The most reliable data were obtained in \cite{BPST06a}
using the Monte Carlo technique: $T_c = 0.16 \mu^{id}$. This result is in
good agreement with experiment. It is noteworthy that the transition
temperature is relatively low, and hence at temperatures somewhat higher
than $T_c$, we face an interesting research object - a degenerate normal
Fermi liquid in the unitary regime.

At finite temperatures, the equation of state cannot be written proceeding
from only the dimensionality considerations. But these considerations lead
to important similarity relations. For instance, the chemical potential is
of the form $\mu (n,T)=\mu ^{id}(n)f_{\mu }[T/\mu ^{id}(n)]$; the entropy
per atom can be written as $s(n,T)=f_{S}[T/\mu ^{id}(n)]$. The last relation
implies that under an adiabatic density variation, the temperature varies as 
$T\propto n^{2/3}$, as in an ideal monatomic gas.

For a fluid in the trap, these formulas lead to an important integral
relation. Following the standard derivation of the virial theorem, it can be
shown that 
\begin{eqnarray}
2U_{pot}=E\ ,
\label{vir}
\end{eqnarray}
where $U_{pot}$ is the potential energy and $E$ is the total energy, i.e.,
the sum of potential, internal, and hydrodynamic kinetic energies \cite%
{Thomas05}. As indicated above, $U_{pot}$ can be calculated directly from
the measured density distribution. The total energy may be changed in a
controllable way. For this, the trap potential was switched off for some
`heating time' $t_{heat}$. During this period, the liquid was free to
expand. The sum of the kinetic and internal energies was conserved in the
process. Then, the trap was turned on again and the system came to
equilibrium, and its potential energy, which was measured anew, turned out
to be higher. This ingenious method enabled the authors of [21] to verify
relation (7) with high precision and thus confirm the similarity laws
formulated above.

The total entropy $S=\int n(\mathbf{x})s(\mathbf{x})d\mathbf{x}$ of the
system as a function of its energy $E$ was measured in a similar experiment
in \cite{Luo06}. In the experiment, the energy was varied and measured as
described above; to measure the entropy, the magnetic field was
adiabatically increased, taking the system away from resonance, where the
interaction was insignificant. Measurements of the cloud dimension enabled
calculating the entropy from the formulas for an ideal Fermi gas, which, due
to the adiabaticity of the process, was equal to the entropy of the liquid
before the increase in the magnetic field. It is noteworthy that the
derivative $T=dE/dS$ directly yields the absolute temperature of the system.
I believe that the capability of measuring the absolute temperature in the
nanokelvin domain is a wonderful achievement by itself. Another way of
measuring the absolute temperature is described below.

The aforesaid leaves no room for doubt that the theoretical notions about
the properties of a `unitary' superfluid liquid are amply borne out by
experiments. I believe, however, that the significance of the issue calls
for high-precision verification. Such a possibility does exist. For this,
the fluid should be placed in a trap that is harmonic and isotropic with a
high degree of accuracy. Then, we can state with certainty that the
spherically symmetric cloud pulsations are precisely equal to $2\omega_{h}$
in frequency, where $\omega_{h}$ is the eigenfrequency of the trap, and do
not attenuate \cite{Castin04}. This theorem is valid both below and above
the superfluid transition point and applies to oscillations of arbitrary
amplitude. It is a corollary of the hidden symmetry of the system at the
unitarity point. (A similar situation occurs for oscillations of a dilute
Bose gas in a cylindrical trap \cite{Pitaevskii96,Kagan96}.) The absence of
damping signifies that the second viscosity $\zeta$ of the fluid is equal to
zero above the transition point. Of the three second viscosity coefficients
introduced by Khalatnikov \cite{Khalatnikov71}, $\zeta_1$ and $\zeta_2$ turn
out to be zero in the superfluid phase \cite{Son07}.

So far, we have dealt with experiments in which the numbers of atoms in two
spin states were equal. Recently, active work commenced to study polarized
systems in which the number of atoms in one spin state (we conventionally
speak of `spin-up' atoms) is greater than in the other state. This question
had already been discussed for superconductors. In \cite{LO64} and \cite%
{FF64}, the existence of spatially inhomogeneous phases (LOFF phases) was
predicted, in which the superconducting gap is a periodic function of
coordinates. In superconductors, the population difference of the spin
states may exist in ferromagnetic bodies or may be induced by an external
magnetic field. In both cases, the magnetic field affects the orbital motion
and destroys superconductivity.

In our neutral dilute systems, the spin relaxation time is quite long and
the numbers of atoms in different states are practically arbitrary
parameters, determined by the initial conditions. Theoretical calculations
 \cite{Lobo06} and the experiment  \cite{Kett07} show that the liquid
in a trap at $T = 0$ near the unitarity point breaks up into three phases.
At the center is the superfluid phase with equal numbers of `spin-up' and
`spin-down' atoms. It is surrounded by the partially polarized normal phase
with unequal densities of the atoms of different polarization. At the
periphery is the completely polarized phase, which consists of only the
atoms of excess polarization. In this case, the existence of LOFF-type
phases in some parameter value ranges is not ruled out.

The measurement data are presented in Fig. 6. The system with $N_{\uparrow
}=5.9\times 10^{6},T/E_{F}=0.03$ and the spin state population ratio $%
N_{\downarrow }/N_{\uparrow }=0.39$ was investigated. Figure 6a shows the
phase-contrast image of the two-dimensional polarization distribution (the column
density) $\delta n_{a}(x,z)\equiv \int dy[n_{\uparrow }(\mathbf{r}%
)-n_{\downarrow }(\mathbf{r})]$ and Fig. 6b shows the `weighted'
distribution $\delta n_{b}(x,z)\equiv \int dy[0.76n_{\uparrow }(\mathbf{r}%
)-1.43n_{\downarrow }(\mathbf{r})]$, which gives a higher-contrast picture.
Figure 6c shows the curves $\delta n_{a}(0,z)$ and $\delta n_{b}(0,z)$, and
Figs 6d and 6e are the plots of integrated linear densities $\delta
n_{a}(z)\equiv \int dxn_{a}(x,z)$ and $\delta n_{a}(x)\equiv \int
dzn_{a}(x,z)$.

\begin{figure}[t]
\includegraphics[width=7cm,height=7cm]{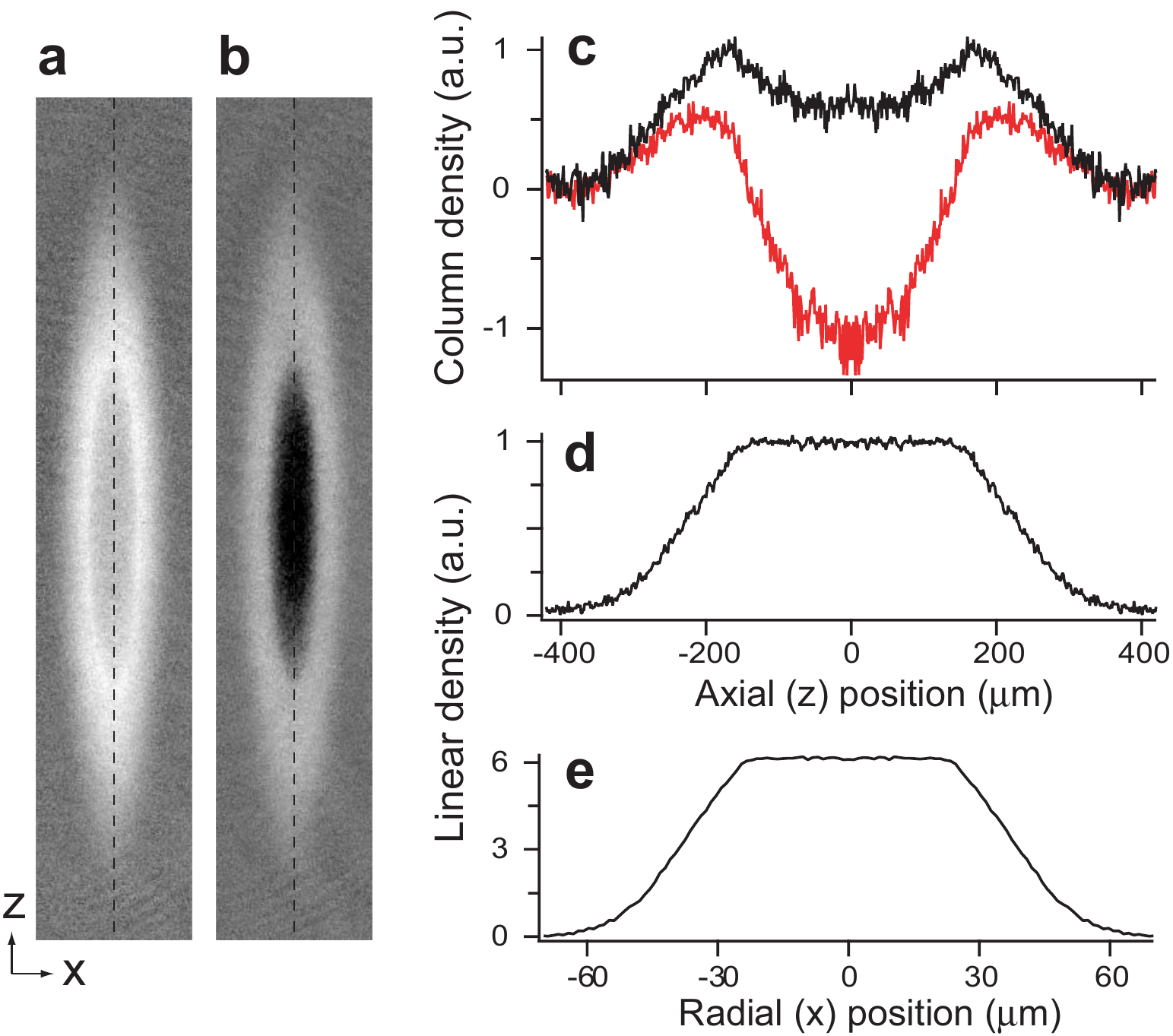}
\caption{Polarization distribution in a system with unequal spin state
populations (from [31]). (a) Phase-contrast image of the two-dimensional
polarization distribution $\delta n_a (x,z)$. (b) Weighted
polarization distribution pattern $\delta n_b (x,z)$. (c) Plots of
the functions $\protect\delta n_a (0,z)$ (the upper curve) and $%
\delta n_b (0,z)$ (the lower curve). (d) Linear polarization density $%
\delta n_a (z)$ along the z axis. (e) Linear polarization density $%
\delta n_a (x)$ along the x axis.}
\label{fig5}
\end{figure}

I emphasize that the measurements were made \textit{in situ}, i.e. in the
trap itself, without prior expansion of the fluid. Processing the measured
two-dimensional distribution by the Abel transform enabled reconstructing
the three-dimensional polarization distribution and confirmed the
three-phase fluid structure. Measurements at different temperatures were
also made. Worthy of note in this connection is the special role played by
the completely polarized phase. Because slow fermions with parallel spins
do not interact with each other, this phase is an ideal Fermi gas. By
measuring the density distribution of this phase and fitting it to formulas
for the ideal gas, it is possible to determine the thermodynamic temperature
of the system. The polarized phase plays the role of an ideal-gas
thermometer contacting with other phases. It is significant in this case
that the fermions of the polarized phase interact with the fermions of other
phases, which ensures thermodynamic equilibrium. These temperature
measurements permitted verifying the transition temperature calculated in
Ref. \cite{BPST06a}. The results under discussion are at some variance with
the findings in \cite{Hulet06}, where a smaller number of atoms was
considered. Conceivably, the surface tension at the phase boundaries plays a
role under these conditions.

I mention several interesting possibilities for future investigations. One
of them involves employing two types of fermions of different masses for
which the Feshbach resonance exists \cite{Petrov03}. Theory predicts
unconventional properties for a superfluid liquid formed as a result of
Cooper pairing of the fermions of different masses.

Another possibility is related to the vortex-free rotation of a Fermi liquid 
\cite{Stringari07}. The vortex lattice shown in Fig. 4 is formed due to a
strong fluid perturbation by the rotating mixers. If a trap asymmetric about
the axis is simply set in rotation, there are grounds to believe that
vortexes would be formed only for a high rotation rate, when the fluid shape
becomes unstable. At lower rotation rates, the fluid would break up into two
phases. The center of the weakly deformed trap would be occupied by the
superfluid liquid at rest, while the normal phase of the liquid would rotate
in the usual way at the periphery. The existence of the normal phase at
absolute zero kept by rotation from transiting into the superfluid state
raises difficult theoretical issues.

A very rich area of research opens up when the fluid is placed in a periodic
lattice produced by counterpropagating laser beams (see the author's review
Ref. \cite{Pitaevskii06}). This research in the unitary domain is still in
its infancy.

We see that the investigations of a near-resonance Fermi gas in a trap have
opened up entirely new theoretical and experimental opportunities in
condensed matter physics, reflecting the modern trend. Work to an increasing
extent is shifting to the investigation of specially fabricated objects that
do not exist in nature and have surprising new properties. In view of this,
I believe, no exhaustion of our realm of physics is to be expected in the
foreseeable future.

I express my appreciation to S. Stringari for discussions, to J. Thomas for
providing the original of Fig. 2, and to R. Grimm for providing the original
of Fig. 5.

\end{document}